 \documentclass[12pt,epsf]{article}
 \usepackage[dvips]{graphicx}
 \usepackage{verbatim,graphics,graphicx,color,slashed,amsmath}
 \usepackage{bm}
 \usepackage{cite}
 \usepackage{amssymb}

\begin{document}
\thispagestyle{empty}
\vspace*{-15mm}
\baselineskip 1pt
\begin{flushright}
\begin{tabular}{l}
{\bf OCHA-PP-357}\\
\end{tabular}
\end{flushright}
\baselineskip 24pt
\vglue 10mm

\vspace{15mm}
\begin{center}
{\Large\bf
Search for Vector-mediated Dark Matter\\at the LHC with Forward Proton Tagging\\
}
\vspace{7mm}

\baselineskip 18pt
{\bf Gi-Chol Cho ${}^{1*}$, Kimiko Yamashita${}^{2, 3\dagger}$, Miki Yonemura${}^{4, 5\ddagger}$}
\vspace{2mm}

{\it
${}^{1}$Department of Physics, Ochanomizu University, Tokyo 112-8610, Japan \\
${}^{2}$Institute of High Energy Physics, Chinese Academy of Sciences, Beijing 100049, China\\
${}^{3}$Department of Physics, National Tsing Hua University, Hsinchu, Taiwan 300\\
${}^{4}$Graduate School of Humanities and Sciences, Ochanomizu University, Tokyo 112-8610, Japan\\
${}^{5}$Program for Leading Graduate Schools, Ochanomizu University, Tokyo 112-8610, Japan
\newline \newline
${}^{*}$cho.gichol@ocha.ac.jp,
${}^{\dagger}$kimiko@ihep.ac.cn,
${}^{\ddagger}$yonemura@hep.phys.ocha.ac.jp}\\
\vspace{10mm}
\end{center}
\begin{center}
\begin{minipage}{14cm}
\baselineskip 16pt
\noindent
\begin{abstract}
We investigate the production of fermionic dark matter $\chi$ via $pp \to p\gamma p \to p j \chi \bar{\chi}X$ mediated by a leptophobic spin-1 particle,
where one of the protons remains intact and is tagged by forward proton detectors.
We find that the masses of $\chi$ and the mediator $Z'$ are severely constrained when $Z'$ interacts with $\chi$ and quarks through the vector couplings.
We show that dark matter searches in this production channel
are sensitive to a mediator mass $m_{Z'} \lesssim 1.4~\mathrm{TeV}$
at 14 TeV at the LHC with an integrated luminosity $L_{\rm{int}} = 3000~\rm{fb}^{-1}$.
The lower mass bound on the dark matter is
$m_\chi \simeq 550~\mathrm{GeV}$
at the mediator mass $m_{Z'}=1.2~\mathrm{TeV}$.

\end{abstract}
\end{minipage}
\end{center}

\baselineskip 18pt
\def\thefootnote{\fnsymbol{footnote}}
\setcounter{footnote}{0}

\newpage

\section{Introduction}
The existence of dark matter (DM) motivates us to explore physics beyond the Standard Model (SM).
Although there are lots of new physics models that explain the origin of DM,
simplified models of dark matter have been adopted as benchmark scenarios to study the DM search strategies at the LHC~\cite{Alves:2011wf}.
Following the recommendations for conducting the systematic DM searches by the LHC Dark Matter Working Group~\cite{Abdallah:2015ter,Abercrombie:2015wmb,Boveia:2016mrp,Albert:2017onk,Abe:2018bpo},
constraints on simplified DM models based on the LHC Run-I and Run-II data have been studied for spin-0~\cite{Khachatryan:2016mdm,CMS:2016pod,Sirunyan:2017jix,Sirunyan:2017qfc,
Sirunyan:2017xgm,Sirunyan:2017leh,Aaboud:2019yqu,Aaboud:2017aeu,Aaboud:2017rzf},
spin-1~\cite{Khachatryan:2016mdm,Aaboud:2016uro,Aaboud:2016tnv,Aaboud:2016qgg,Aaboud:2016obm,Sirunyan:2016iap,CMS:2016pod,CMS:2016hmx,CMS:2016fnh,Sirunyan:2017hci,Aaboud:2017buf,Aaboud:2017phn,Sirunyan:2017jix,Aaboud:2019yqu},
and spin-2 mediators~\cite{Lee:2013bua,Lee:2014caa,Han:2015cty,Dillon:2016tqp,Kraml:2017atm,Carrillo-Monteverde:2018phy,
Carrillo-Monteverde:2018jcs}.

%
At the LHC, in addition to the central detectors, forward proton detectors have been installed, such as
the ATLAS Forward Proton detector (AFP)~\cite{Adamczyk:2015cjy} and CMS-TOTEM Precision Proton Spectrometer (CT-PPS)~\cite{Albrow:2014lrm}.
These forward proton detectors enable us to study processes with photons in the initial state that are induced from initial protons.
Although such non-QCD processes might give us new strategies to look for DM at the LHC, the feasibility of searching for DM candidates of simplified DM models using the forward proton detectors has not been fully examined yet.
Neutralino searches via the forward proton detectors were studied in Refs.~\cite{Schul:2008sr, Harland-Lang:2018hmi},
and in Ref.~\cite{Harland-Lang:2018hmi}, a detailed feasibility study
was performed regarding neutralino searches with proton tagging in compressed mass scenarios of
the supersymmetric Standard Model.

%
%
%
The forward proton detectors  AFP and CT-PPS are installed symmetrically at about 210~m
from the interaction point~\cite{Erland:2019ghv,Albrow:2015ois}.
These forward proton detectors detect intact protons with the momentum fraction loss
\begin{align}
\xi \equiv \frac{|\vec{p}|-|\vec{p}^{\, \prime}|}{|\vec{p}|},
\label{eq:xi}
\end{align}
where $\vec{p}$ and $\vec{p}^{\, \prime}$ denote the momentum of an initial proton and a forward proton after elastic photon emission, respectively.
The acceptance $\xi$ of a forward proton detector in ATLAS and CMS is~\cite{Baldenegro:2018hng,}
\begin{align}
0.015 < \xi < 0.15.
\label{eq:xi_acceptance}
\end{align}
Possibilities to search for new physics beyond the SM via photon-photon or photon-proton collisions at the LHC have been discussed in, e.g.,
Refs.~\cite{Ginzburg:1999ej,Khoze:2001xm,Schul:2008sr,Kepka:2008yx,Dougall:2007tt,Chaichian:2009ts,Sahin:2009gq,
Chapon:2009hh,Piotrzkowski:2009sa,Inan:2010af,Atag:2010bh,Goncalves:2010dw,Sahin:2010zr,Sahin:2011yv,
Gupta:2011be,Epele:2012jn,Sahin:2012ry,Sahin:2013qoa,Koksal:2014hba,Lebiedowicz:2013fta,Fichet:2013ola,
Fichet:2013gsa,Sun:2013ria,Sun:2014qoa,Sun:2014qba,Sun:2014ppa,Sahin:2014dua,Inan:2014mua,
Senol:2014vta,Alva:2014gxa,Fichet:2014uka,Fayazbakhsh:2015xba,Ruiz:2015gsa,Fichet:2015nia,Cho:2015dha,
Fichet:2016clq,Fichet:2015vvy,Koksal:2017nmy,Baldenegro:2018hng,Inan:2018jza,Sahin:2019doi,Heinemeyer:2007tu,Tasevsky:2014cpa, Das:2015toa, Harland-Lang:2018hmi}.

%
In this article, we study the feasibility of looking for signatures or constraints on the simplified DM model with fermionic DM
and a spin-1 mediator using the forward proton detectors.
%
%
The fermionic DM $\chi$ does not interact with the SM particles directly,
while a spin-1 mediator $Z'$ couples to both the DM and SM particles.
Since a massive vector boson is strongly constrained at the LHC through the resonance search in the dilepton channel~\cite{Aaboud:2017buh,Sirunyan:2018exx},
we restrict ourselves to consider the case of a leptophobic vector mediator.

%
%
\begin{figure}[t]\center
 \includegraphics[width=0.495\textwidth,clip]{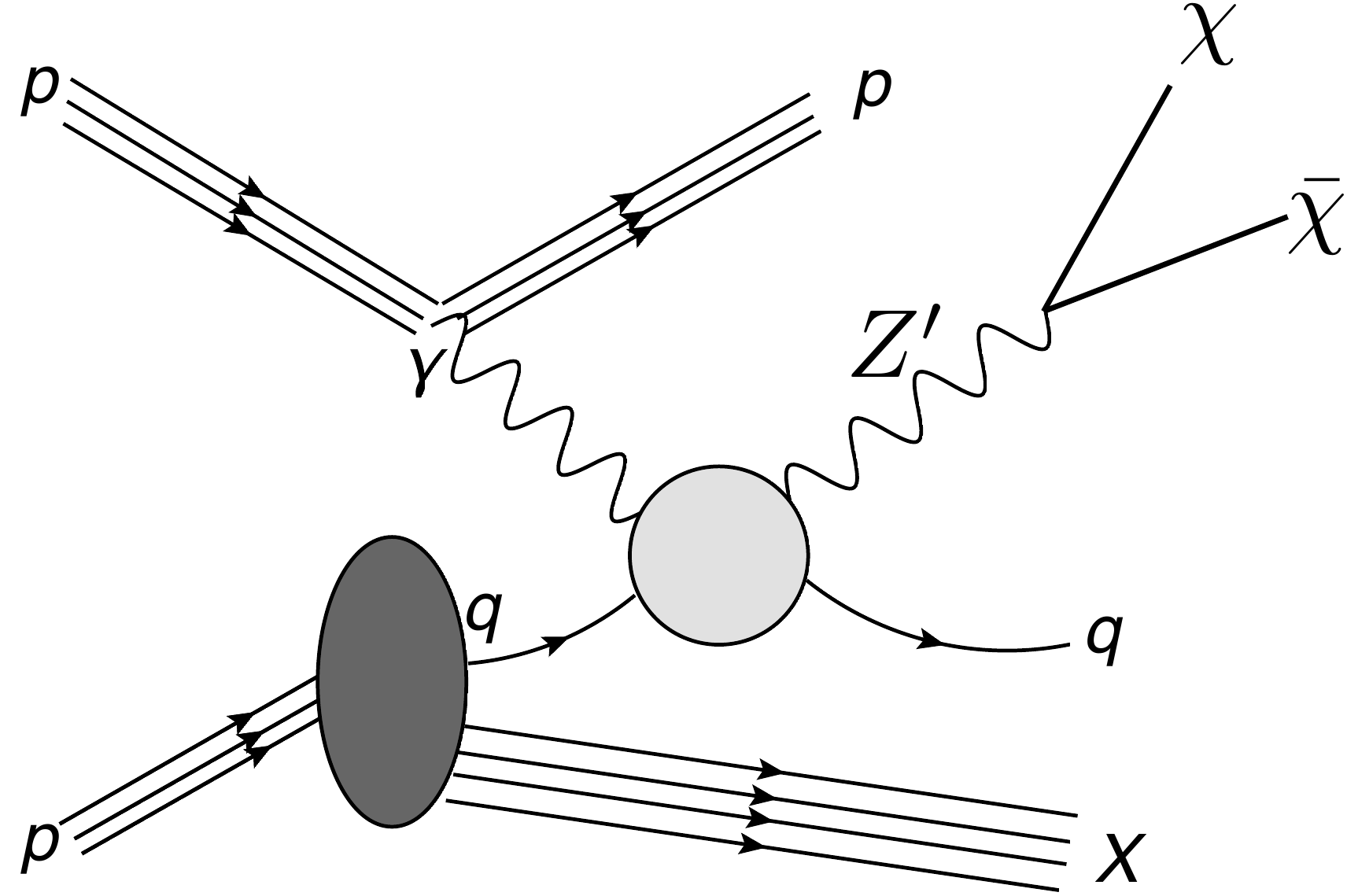}
  \includegraphics[width=0.495\textwidth,clip]{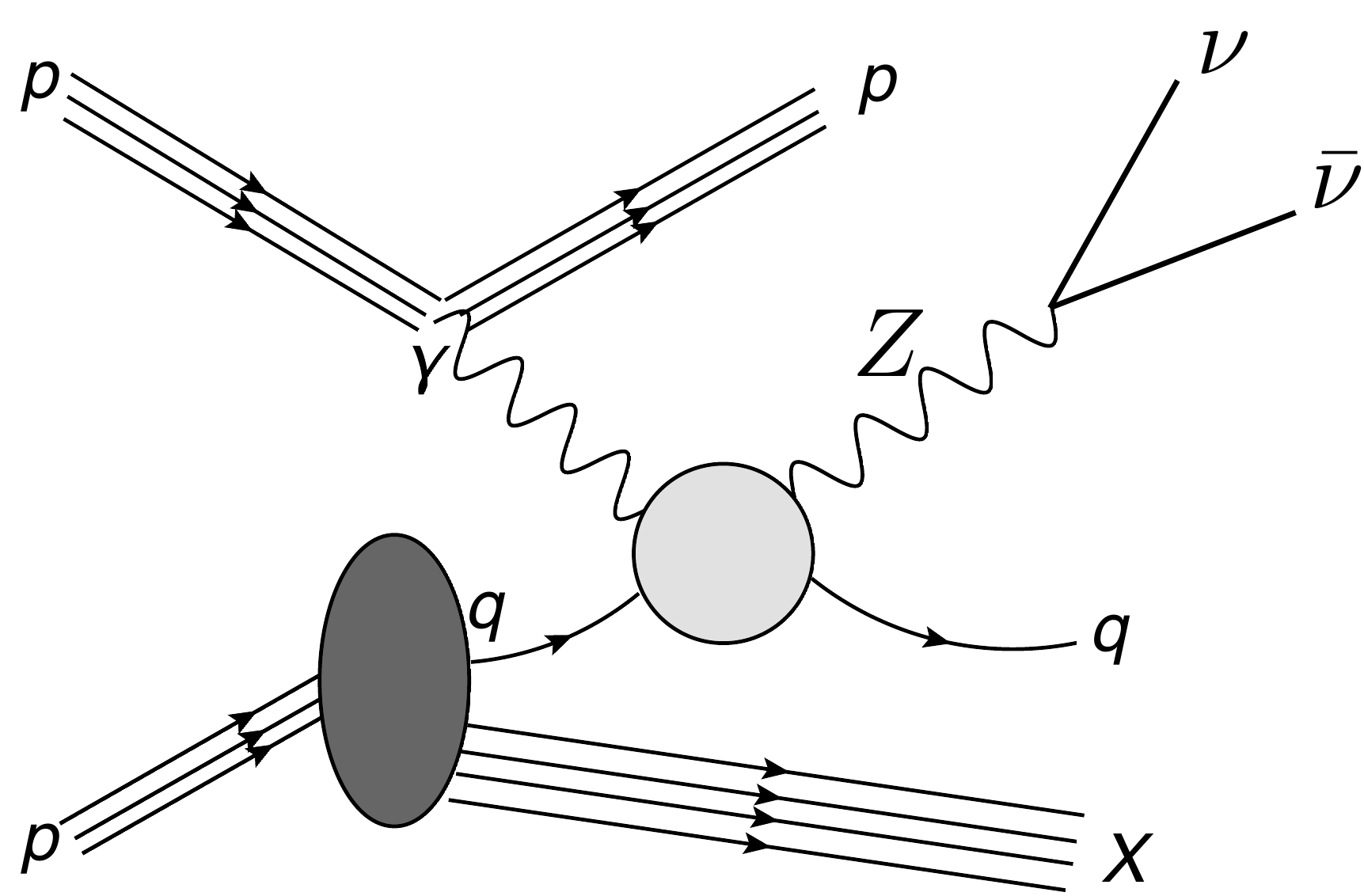}
\caption{
The DM pair production process
(left) and the SM background process (right) at the LHC with proton tagging at the forward proton detector.
%
%
}
\label{fig:pp_aq_Yj}
\end{figure}

%
The production process of the DM in our study is
\begin{align}
pp\to p\gamma p\to p j \chi \bar{\chi} X,
\label{eq:prod_proc}
\end{align}
where the DM $\chi$ is the Dirac fermion and
$j=u,d,c,s,b$ (and their antiparticles).
The main background process is
\begin{align}
pp\to p\gamma p\to p j \nu \bar{\nu} X,
\label{eq:bg_proc}
\end{align}
where $\nu \bar{\nu}$ is summed over three flavors of neutrinos.
We depict the DM production process and the SM background process in Fig.~\ref{fig:pp_aq_Yj}.
%
%
%
In both signal and background processes,
a quasireal photon $\gamma$ is emitted from a proton and scattered with
a parton in the proton coming from the opposite direction.
The proton that emits the quasireal photon does not break up into partons, but rather loses its momentum and is finally detected by the forward proton detector.
%
%
The momentum fraction loss of the intact proton is estimated by $\xi$ in Eq.~(\ref{eq:xi}).
As will be shown later, the SM background events could be sizably reduced by appropriate cuts on $\xi$.
%
%

%
The DM production process (\ref{eq:prod_proc}) via forward proton detectors
was investigated in Ref.~\cite{Sun:2014ppa} based on the effective field theory (EFT) framework.
In the EFT approach, pair production of DM is described in terms of contact interaction operators, so it is a good approximation only when the mediator mass is large enough as compared to the energy scale at the LHC.
On the other hand, in simplified DM models, the DM pair is produced by the mediators so that the lower mass region of the mediator can be analyzed.
%
%

%
This article is organized as follows.
In Sec.~\ref{sec:model}, we briefly review
a simplified DM model with a leptophobic vector mediator.
The numerical analyses of the signal and background processes
are given in Sec.~\ref{sec:analysis}.
Constraints on the model parameters are shown in Sec.~\ref{sec:results}.
Section~\ref{sec:summary} is devoted to a summary.
\section{Model}\label{sec:model}
In this section we briefly review the interactions of the DM $\chi$ and leptophobic spin-1 mediator $Z'$ in the simplified DM model~\cite{Backovic:2015soa}.
The interaction Lagrangian of the spin-1 mediator $Z'$ and a fermion $\psi$ is given by
\begin{align}
{\cal L}^{Z'}_{\mathrm{int}}
=
\overline{\psi}\gamma^{\mu}(g^V_\psi +g^A_\psi \gamma_5) \psi {Z'}_{\mu},
\label{eq:lag_dm0}
\end{align}
where $g^V_\psi$ and $g^A_\psi$ denote vector and axial-vector couplings of the mediator $Z'$ for $\psi$, respectively.
Since the mediator $Z'$ is leptophobic in our study, the fermion $\psi$ in
Eq.~(\ref{eq:lag_dm0}) represents the fermionic DM $\chi$ and
quarks $q(=u,d,c,s,b,t)$.
Then, the production process of the DM (\ref{eq:prod_proc}) could be studied quantitatively using the following model parameters:
the dark matter mass $m_\chi$, the mediator mass $m_{Z'}$ and the couplings of fermions $g^V_\psi$ and $g^A_\psi$.
Throughout our study, we consider that
vector and axial-vector couplings of quarks to the mediator--$g^V_q$ and $g^A_q$--are generation independent for simplicity.

%
We study constraints on $m_\chi$ and $m_{Z'}$
from the process (\ref{eq:prod_proc})
based on three reference scenarios for the interactions of the mediator $Z'$:
%
\begin{enumerate}
\renewcommand{\theenumi}{(\roman{enumi})}
\renewcommand{\labelenumi}{(\roman{enumi})}
 \item vector couplings only (``vector scenario"),
       \begin{align}
	g^V_{\chi} = 1.0,~
	g^A_{\chi} = 0.0,~
	g^V_q = 0.25,~
	g^A_q = 0.0;
	\label{eq:lag_dm1}
       \end{align}

\item axial-vector couplings only (``axial-vector scenario"),
      \begin{align}
       g^V_{\chi} = 0.0,~
       g^A_{\chi} = 1.0,~
       g^V_q = 0.0,~
       g^A_q = 0.25;
       \label{eq:lag_dm2}
      \end{align}

\item combination of vector and axial-vector couplings
 (``mixed scenario"),
      \begin{align}
       g^V_{\chi} = \frac{1}{\sqrt{2}},~
       g^A_{\chi} = \frac{1}{\sqrt{2}},~
       g^V_q = \frac{1}{4\sqrt{2}},~
       g^A_q = \frac{1}{4\sqrt{2}},
       \label{eq:lag_dm3}
      \end{align}
\end{enumerate}
where the values of the couplings in Eqs.~(\ref{eq:lag_dm1}) and (\ref{eq:lag_dm2}) are adopted from
benchmark scenarios by the LHC Dark Matter Working Group~\cite{Albert:2017onk}.

\section{Numerical Analysis}\label{sec:analysis}
\begin{figure}[t]
	\hspace{3.5cm}(a)\hspace{7cm}(b)\\
	\vspace{-1cm}
	\center
	\includegraphics[width=0.45\textwidth,clip]{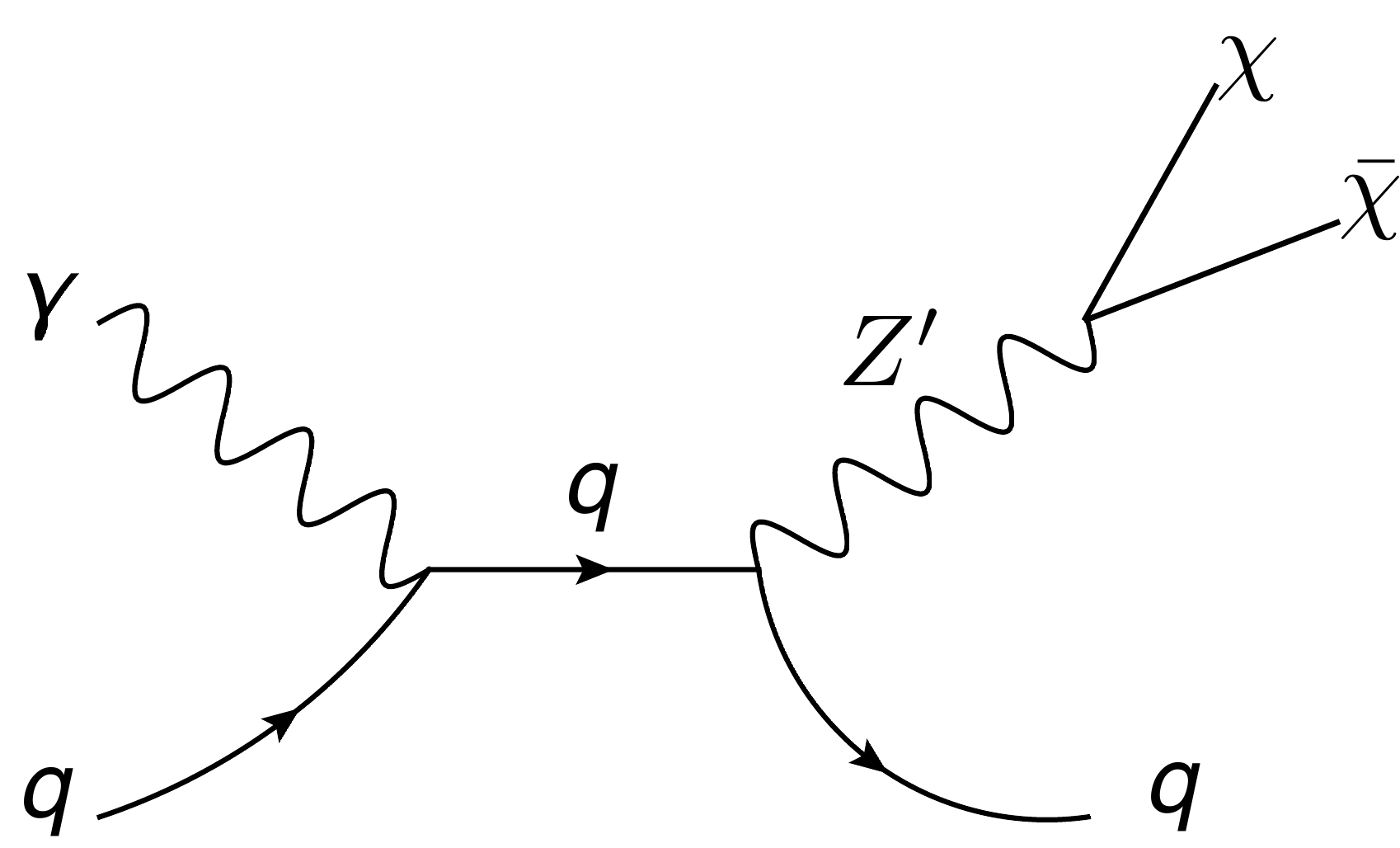}\,
	\includegraphics[width=0.45\textwidth,clip]{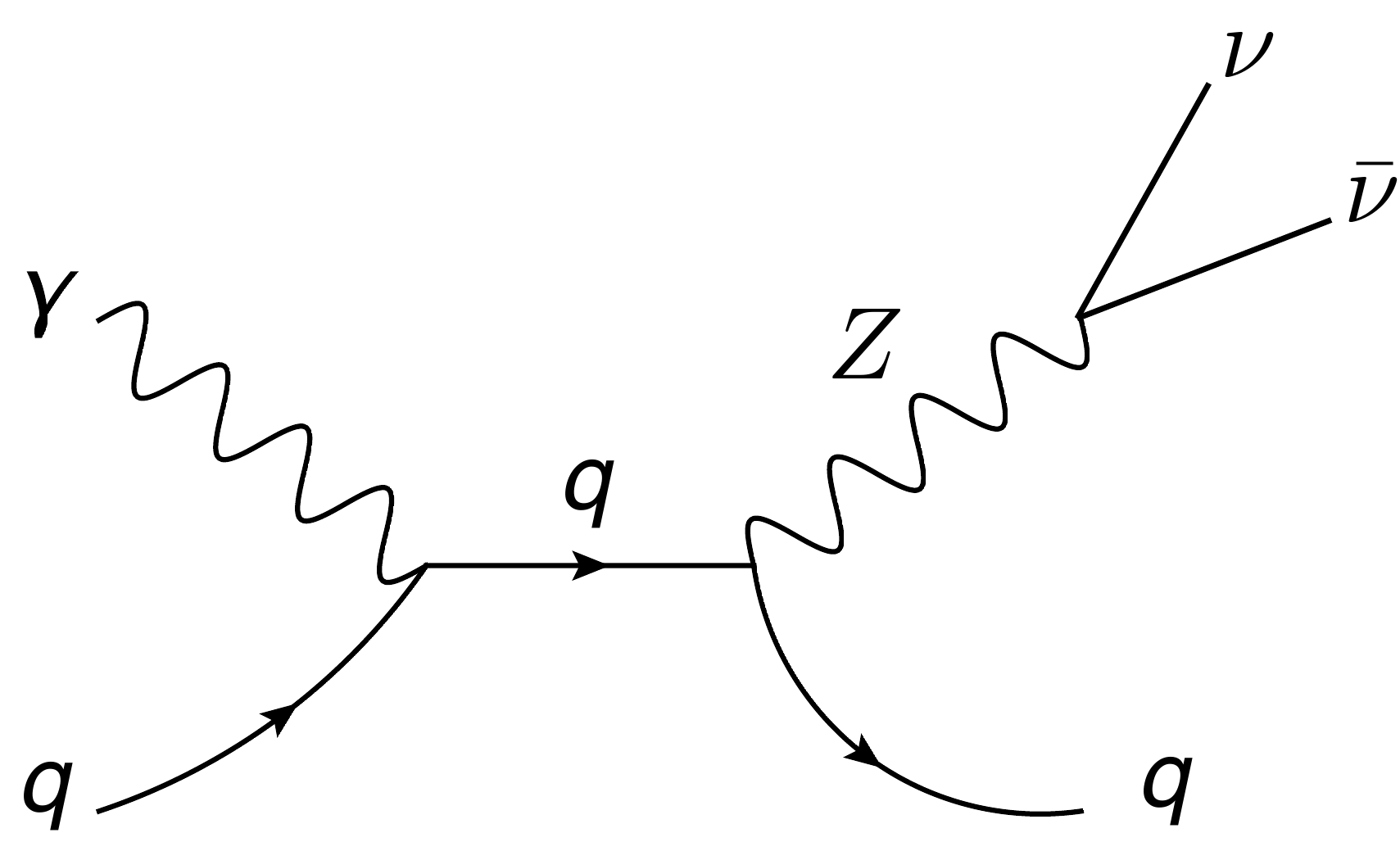}\\
	\includegraphics[width=0.35\textwidth,clip]{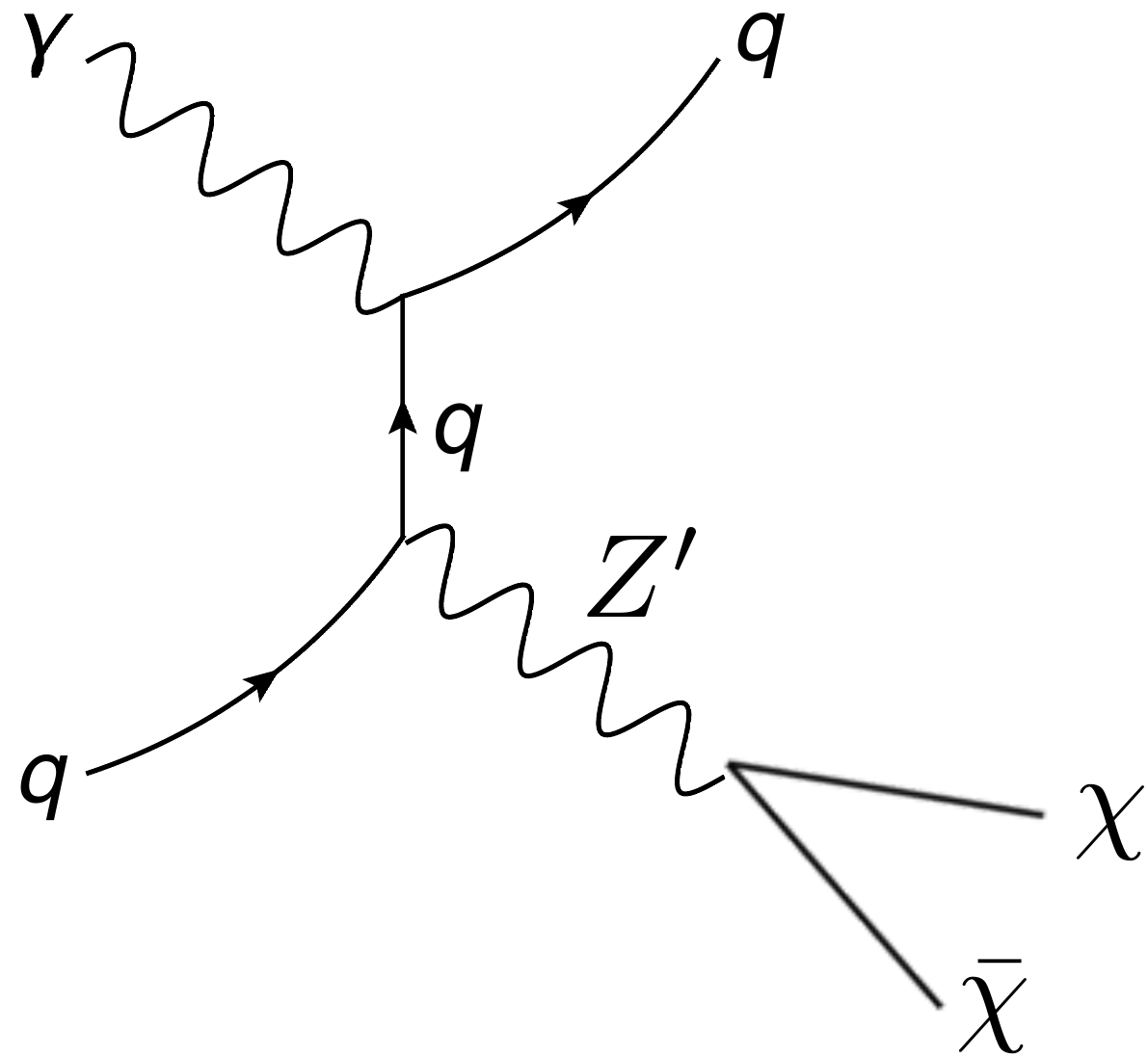}\, \, \, \, \, \, \, \, \,
	\includegraphics[width=0.35\textwidth,clip]{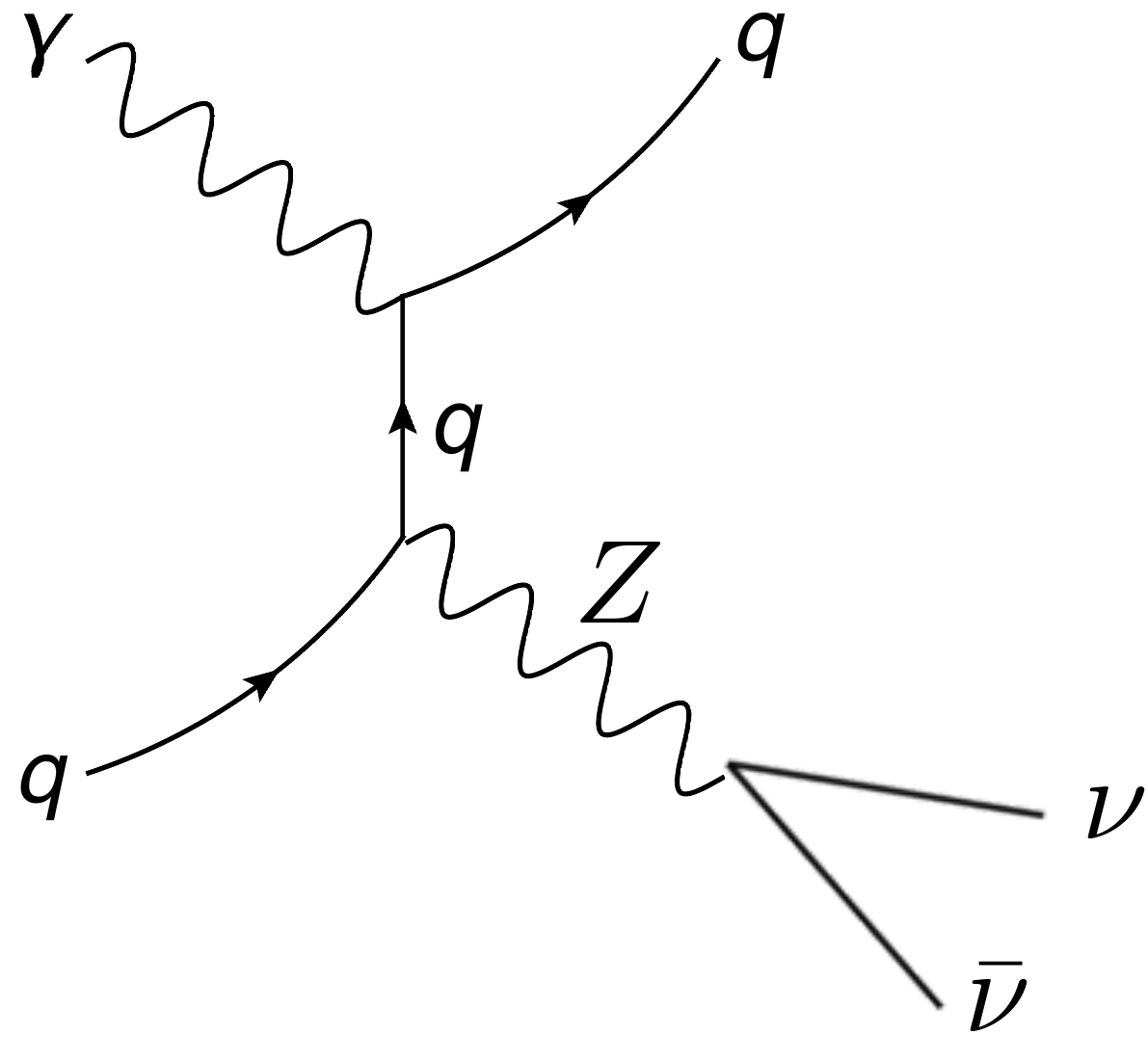}
 \caption{Parton-level Feynman diagrams for the signal (a) and SM background (b).
Here $q=u,d,c,s,b,\bar{u},\bar{d},\bar{c}, \bar{s}$, and $\bar{b}$ and $\nu=\nu_{e}, \nu_{\mu},$ and $\nu_{\tau}$.}
\label{fig:signal_bg}
\end{figure}

We show parton-level Feynman diagrams for the signal process (\ref{eq:prod_proc}) and primary background process (\ref{eq:bg_proc})
in Figs.~\ref{fig:signal_bg}(a) and (b), respectively.
The collider signature of these processes is a jet plus missing energy with the intact proton detected at the forward proton detector.
We assume that $Z'$ promptly decays into the two neutral particles $\chi$ and $\bar{\chi}$.
In fact, the $Z' \to \chi \bar{\chi}$ decay widths are 26, 20, and 23 GeV for
the vector and axial-vector couplings in Eqs.~(\ref{eq:lag_dm1})--(\ref{eq:lag_dm3}), respectively,
with $(m_{Z'}, m_\chi) = (1~\rm{TeV}, 200~\rm{GeV})$.
We also assume that $\chi$ and $\bar{\chi}$  are stable, and do not to decay inside the main LHC detector.
Thanks to the stability of the DM, the final state contains the missing transverse energy $\ensuremath{\slashed{E}_T}$.
We also include the off-shell $Z'$ mode in the computation of the signal events.
The final state of the signal includes an intact proton, one jet, the missing transverse energy $\ensuremath{\slashed{E}_T}$, and $X$, which is the proton remnant.
The proton that emits the quasireal photon does not break up into partons, but rather loses its momentum.
This proton travels with a slightly different angle from the beam because of the magnetic fields at the LHC.
Finally, the proton is detected by the forward proton detector.
One jet and the missing transverse energy $\ensuremath{\slashed{E}_T}$ are measured at the central detector at the LHC.
%

%
%
There is another SM background process $pp \to p\gamma p \to p j \nu \bar{\nu} \nu \bar{\nu} X$.
However, this cross section is about 2000 times smaller than the leading background process
 $pp \to p\gamma p \to p j \nu \bar{\nu} X$, and thus it is quantitatively negligible.

%
We employ {\sc MadGraph5\_aMC@NLO}~\cite{Alwall:2014hca} to generate parton-level events for both the signal and background processes with NNPDF2.3~\cite{Ball:2012cx}. The interactions of the spin-1 mediator and fermions (DM and quarks) are implemented
by using the spin-1 DMsimp model file~\cite{UFO}.
The flux of quasireal photons emitted from a proton via the equivalent photon approximation~\cite{Budnev:1974de} is implemented in {\sc MadGraph5\_aMC@NLO},
in which fully elastic contributions are taken into account.

In the following study, we fix the center-of-mass energy $\sqrt{s}=14~\mathrm{TeV}$
and the integrated luminosity $L_{\rm{int}} = 3000~\mathrm{fb}^{-1}$.
The survival probability of a proton ($S^2$) after photon emission is $S^2=0.7$~\cite{Sun:2014ppa}.
This survival probability depends on some processes
(for diffractive processes, see Refs.~\cite{Sun:2014ppa,Khoze:2001xm,Jones:2013pga}).
In our case the other proton breaks up into partons, and a monojet signal is detected by the central detector.
As minimal event selections,
we impose cuts on the transverse momentum $p_T^j$ and pseudorapidity $\eta^j$ for the jet as
\begin{align}
 p_T^{j}>200~{\rm GeV},\quad |\eta^{j}|<3.0.
\label{eq:min_cuts}
\end{align}
In the parton-level analysis, the cut $p_T^{j}>200~{\rm GeV}$
in Eq.~(\ref{eq:min_cuts}) is equivalent to the selection cut on
the missing transverse energy $\ensuremath{\slashed{E}_T}>200~{\rm GeV}$.
For the missing transverse energy we adopt the definition~\cite{Conte:2012fm}
\begin{align}
 \ensuremath{\slashed{E}_T} = \| \sum_{\text{visible particles}} \vec{p}_T\|,
\label{eq:missing_energy}
\end{align}
because in our parton-level analysis one quark is emitted for one event,
$\ensuremath{\slashed{E}_T} = p^j_T$,
where the jet with $p^j_T$ corresponds to the quark.


%
\begin{figure}[t]\center
 \includegraphics[width=0.496\textwidth,clip]{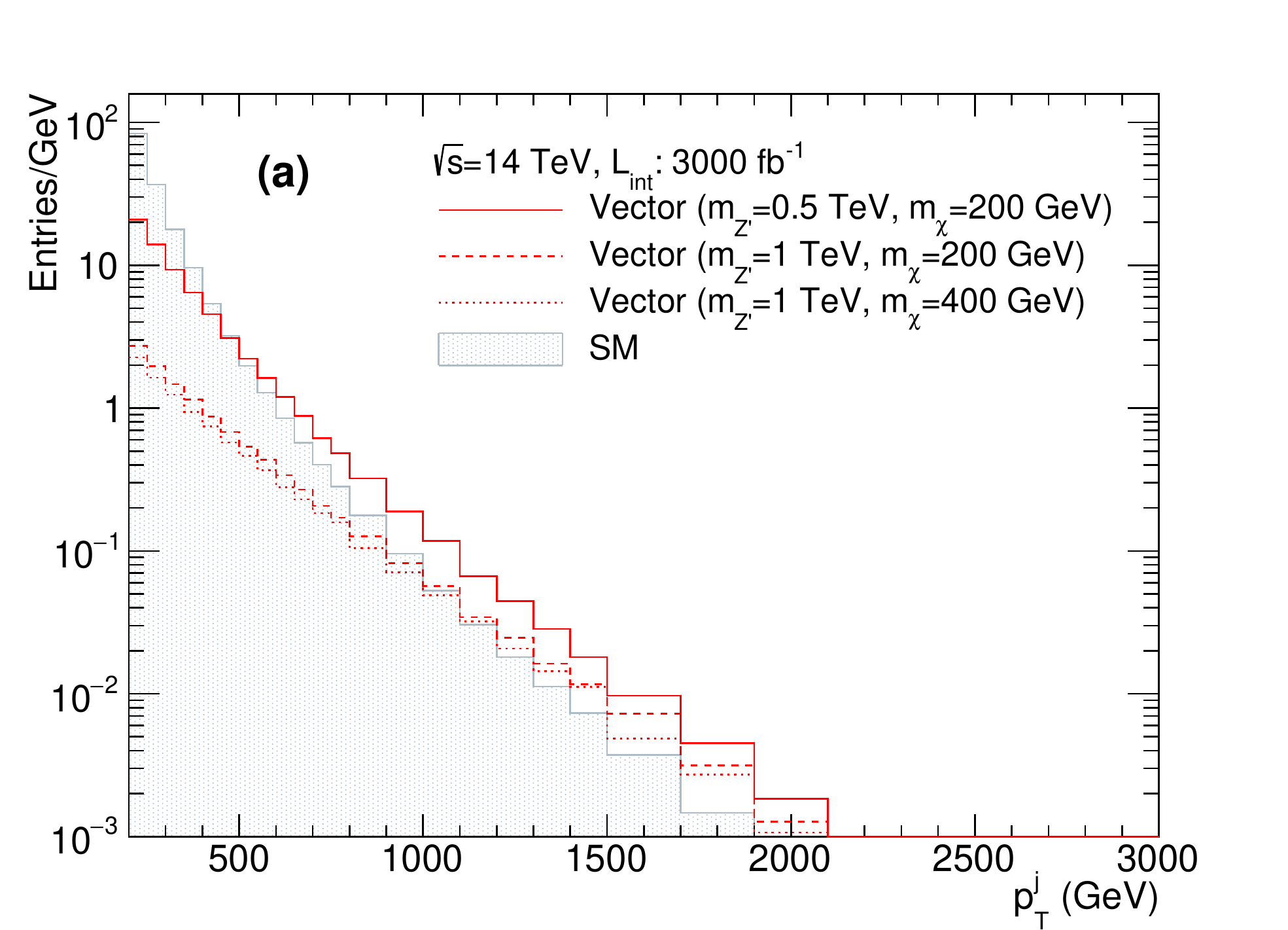}\\
 \includegraphics[width=0.496\textwidth,clip]{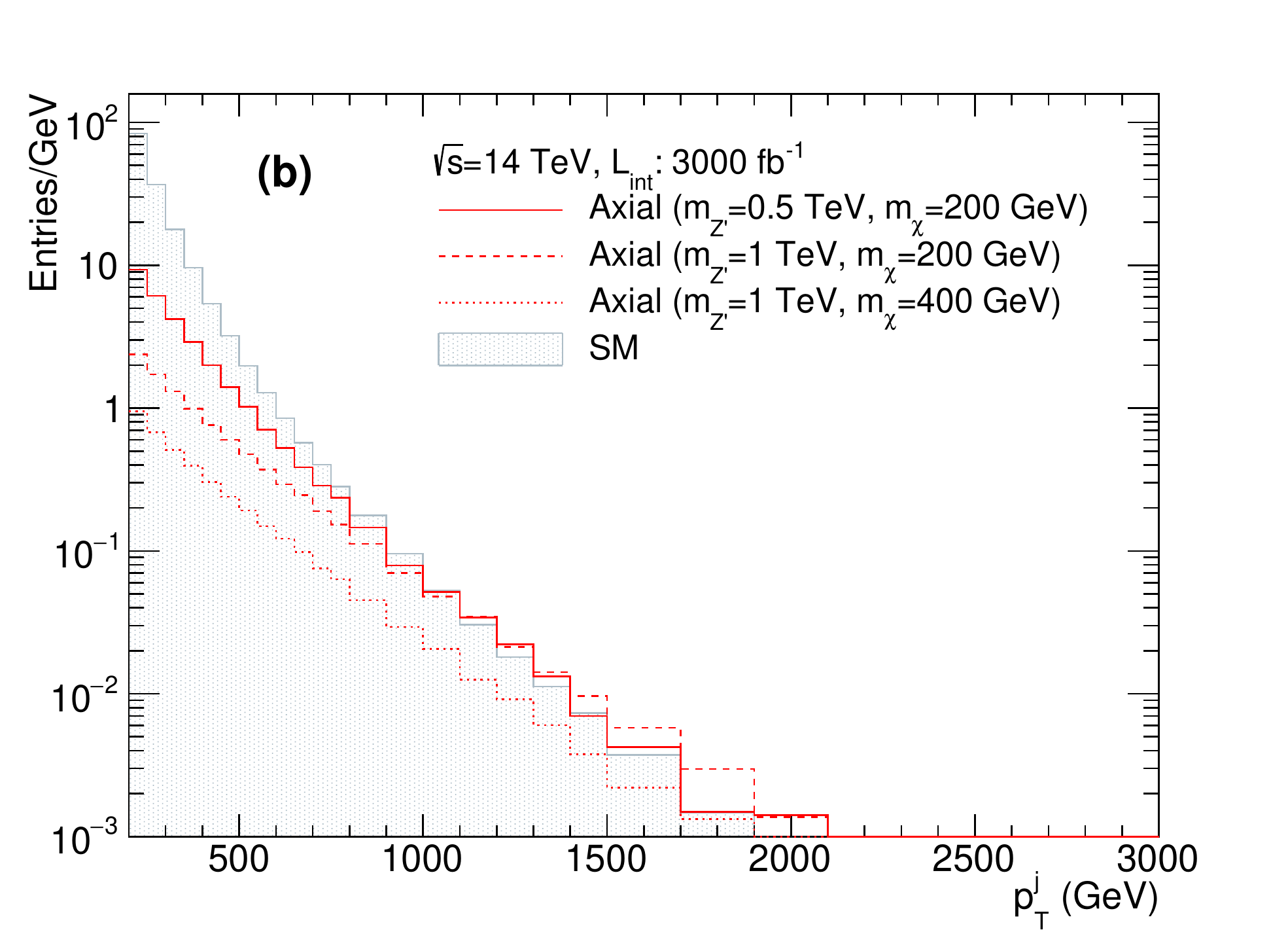}
 \includegraphics[width=0.496\textwidth,clip]{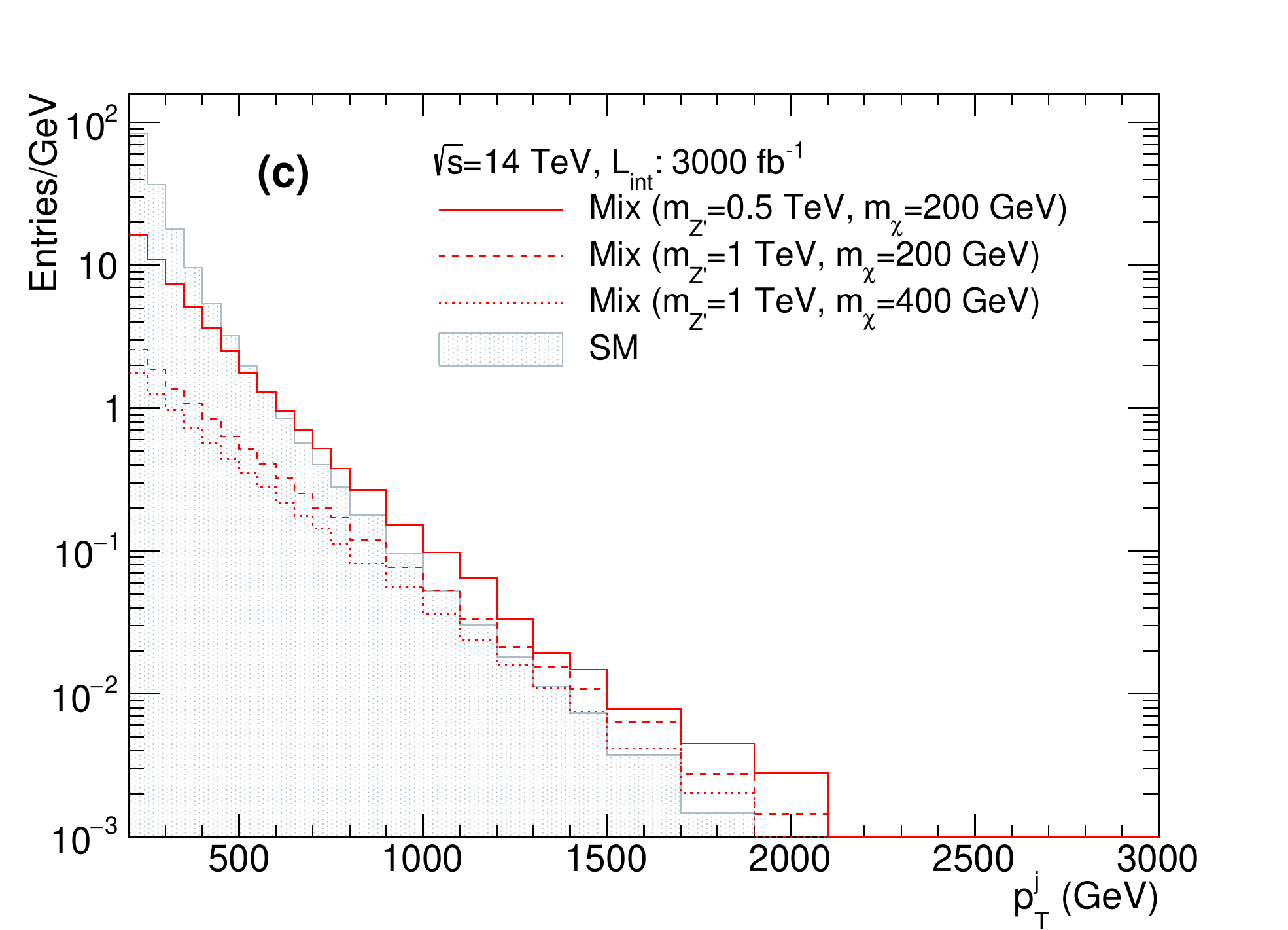}
\caption{
The
$p_T^j$ distributions of $pp\to p\gamma p\to p j \chi \bar{\chi} X$
at $\sqrt{s}=14~\mathrm{TeV}$ and $L_{\mathrm{int}}=3000~\mathrm{fb}^{-1}$ for
(a) the vector scenario,
(b) the axial-vector scenario, and
(c) the mixed scenario.
The vector and axial-vector couplings of $Z'$ for panels (a)--(c) are given in
Eqs.~(\ref{eq:lag_dm1})--(\ref{eq:lag_dm3}).
The three lines in each panel correspond to
$(m_{Z'}, m_\chi) = (500~\mathrm{GeV}, 200~\mathrm{GeV})$ (solid),
$(1~\mathrm{TeV}, 200~\mathrm{GeV})$ (dashed), and
$(1~\mathrm{TeV}, 400~\mathrm{GeV})$ (dotted),  respectively.
The shaded region indicates the distribution of the SM background events.
}
\label{fig:pt}
\end{figure}
In Figs.~\ref{fig:pt} (a)-(c), we show the $p_T^j$ distributions of the signal process  (\ref{eq:prod_proc})
in the simplified DM model with the spin-1 mediator for the three
scenarios (\ref{eq:lag_dm1}), (\ref{eq:lag_dm2}), and (\ref{eq:lag_dm3}),
respectively.
In each figure, the solid, dashed, and dotted lines correspond to
$(m_{Z'}, m_\chi) = (500~\rm{GeV}, 200~\rm{GeV})$,  $(1~\rm{TeV}, 200~\rm{GeV})$, and
$(1~\rm{TeV}, 400~\rm{GeV})$, respectively.
We used the vector and axial-vector couplings for quarks and the DM
in Eqs.~(\ref{eq:lag_dm1}), (\ref{eq:lag_dm2}), and
(\ref{eq:lag_dm3}) for each scenario.
The $p_T^j$ distribution of the SM background process (\ref{eq:bg_proc}) is also shown by the shaded region for comparison in each figure.
No significant difference between the $p^j_T$ distributions of the signal and background events is found after applying the cut $p^j_T > 200$~GeV.

%
\begin{figure}[t!]\center
 \includegraphics[width=0.496\textwidth,clip]{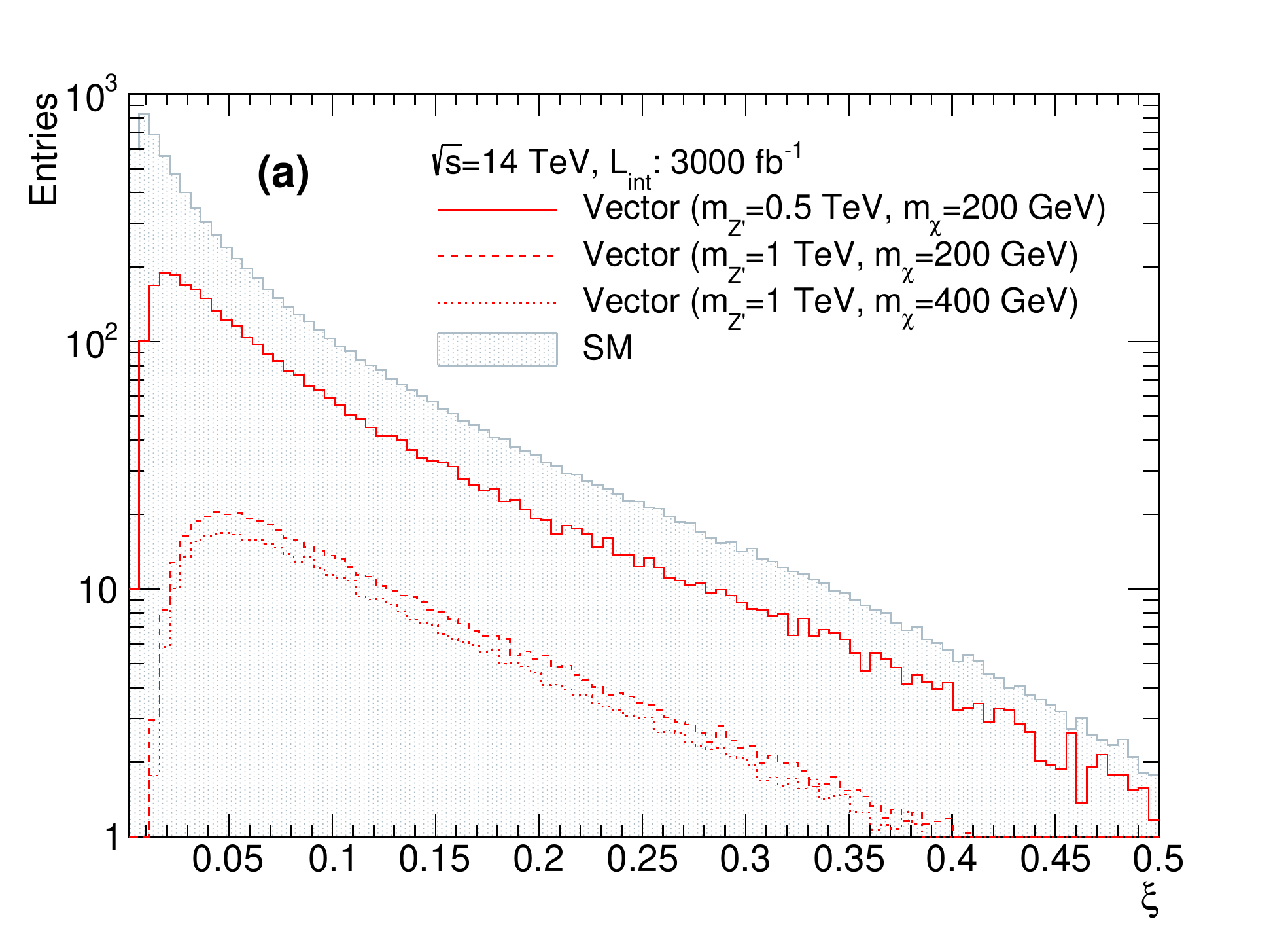}
 \includegraphics[width=0.496\textwidth,clip]{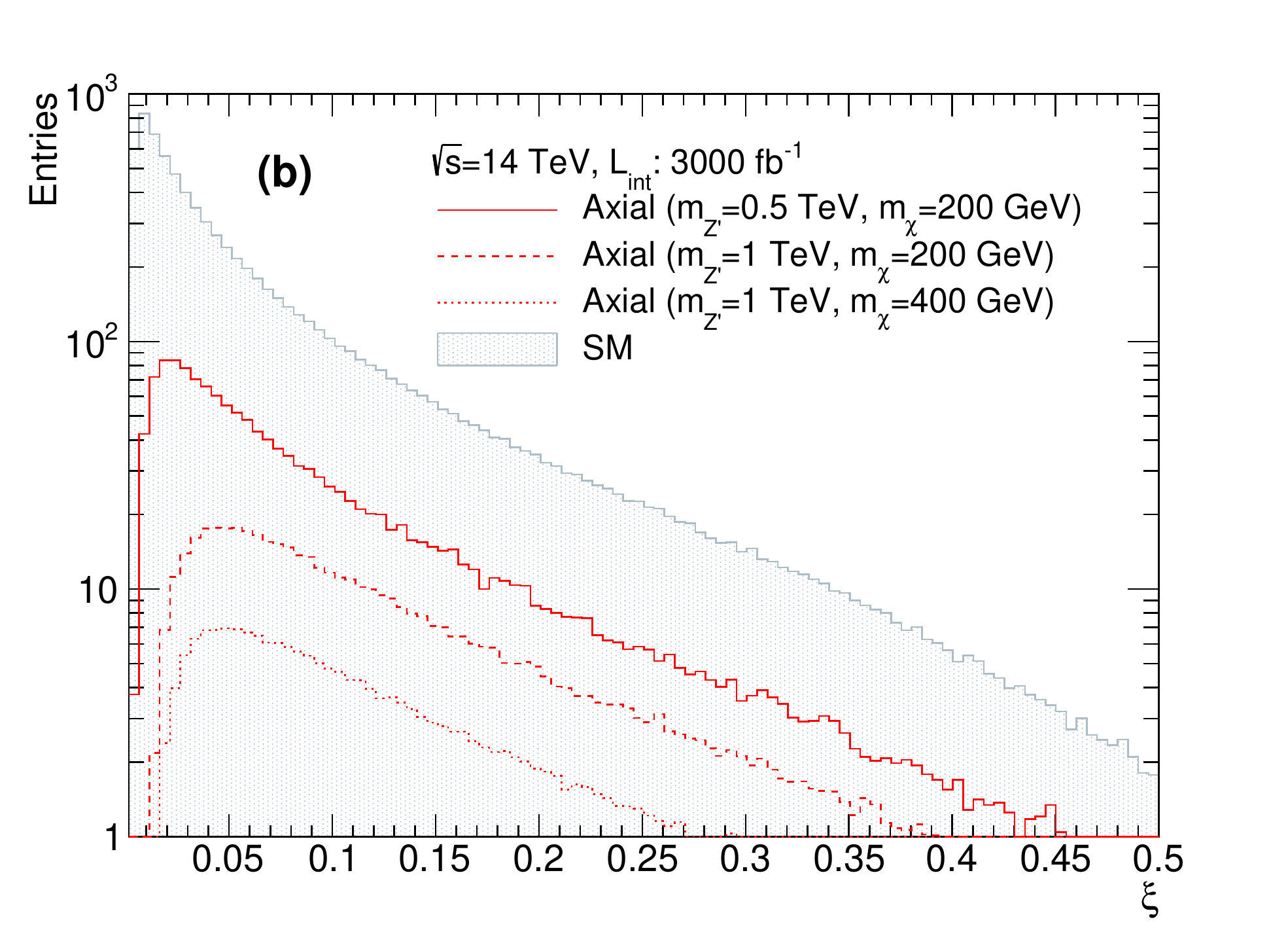}\\
 \includegraphics[width=0.496\textwidth,clip]{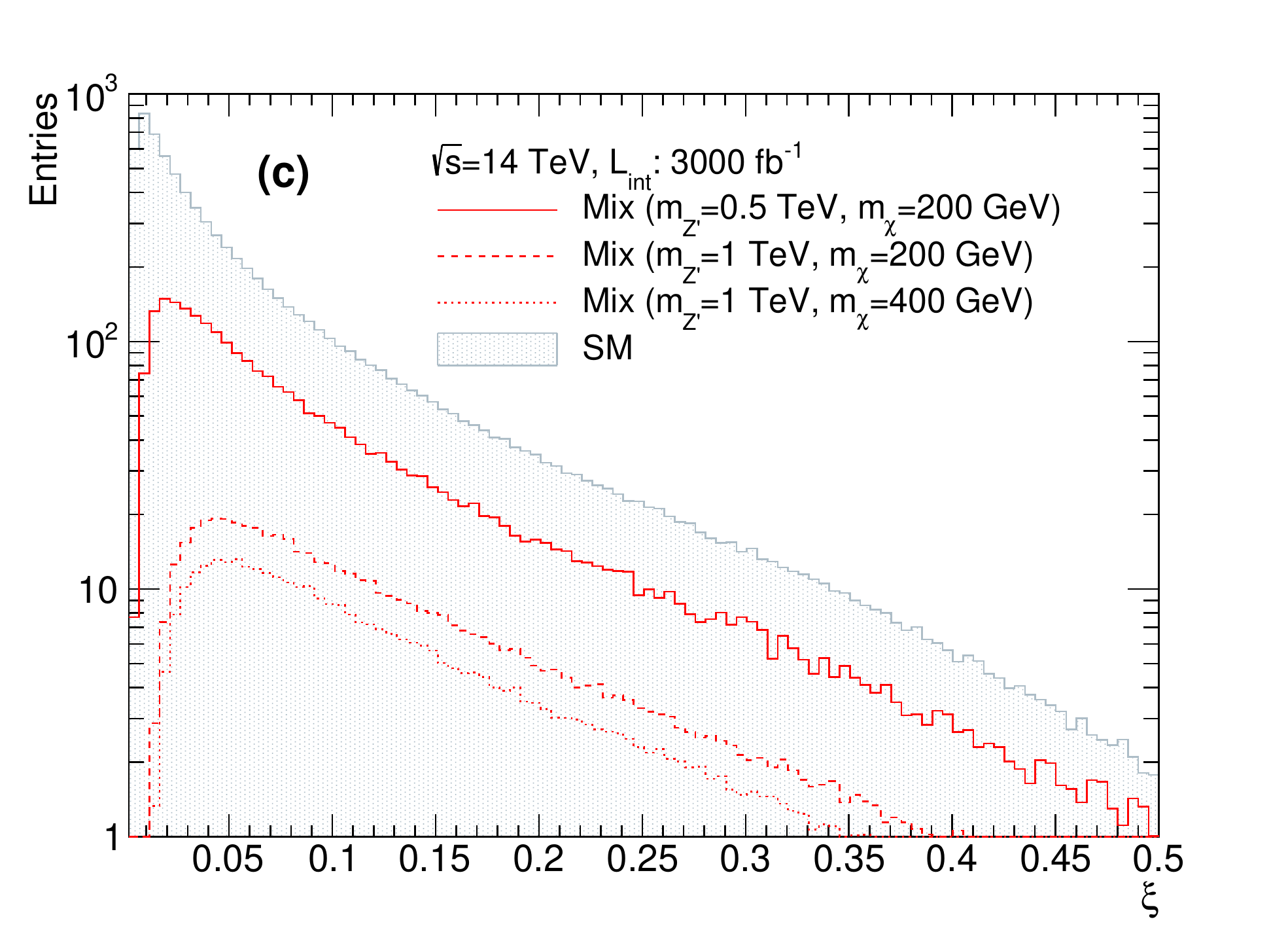}
   \includegraphics[width=0.496\textwidth,clip]{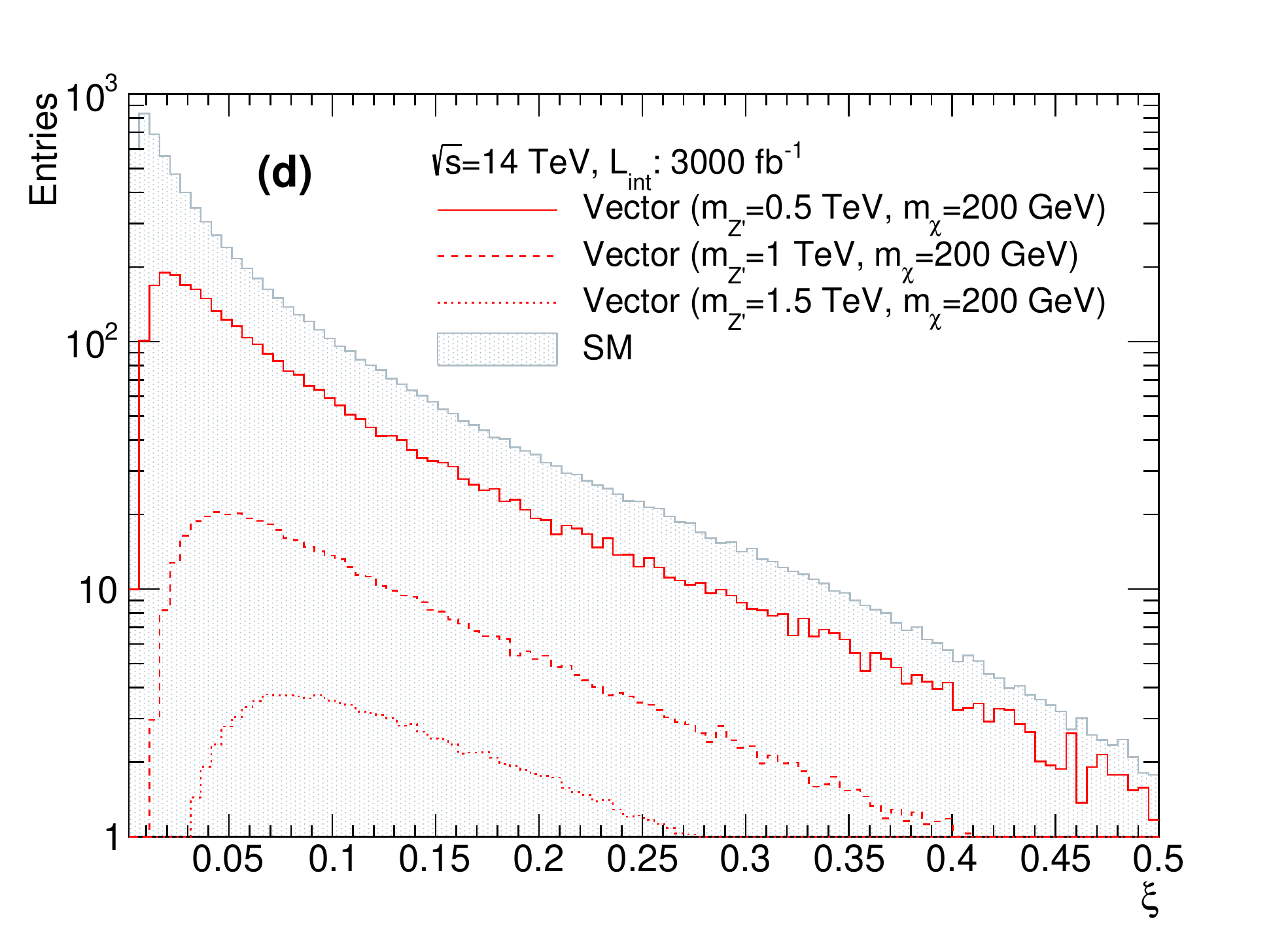}
\caption{
The $\xi$ distribution of the signal process $pp\to p\gamma p\to p j \chi \bar{\chi} X$ at $\sqrt{s}=14~\mathrm{TeV}$ and $L_{\mathrm{int}}=3000~\mathrm{fb}^{-1}$.
The parameter sets in panels (a), (b), and (c) are the same in Figs.~\ref{fig:pt}(a), \ref{fig:pt}(b), and \ref{fig:pt}(c).
In panel (d), the same values of couplings as in panel (a) are used, and
the three lines indicate
$m_{Z'} = 500~\mathrm{GeV}$ (solid), $1~\mathrm{TeV}$ (dashed),
and $1.5~\mathrm{TeV}$ (dotted), respectively, with $m_\chi = 200~\mathrm{GeV}$.
}
\label{fig:xi}
\end{figure}
Next, we show the $\xi$ distributions of the signal and background
processes in Fig.~\ref{fig:xi}.
The shaded region in each figure indicates the SM background.
The three lines in Figs.~\ref{fig:xi}(a)--\ref{fig:xi}(c)
are obtained using the same input values of $m_{Z'}$ and $m_\chi$
as in Fig.~\ref{fig:pt}.
In Fig.~\ref{fig:xi}(d) we compare
the $\xi$ distributions for three different values of the $Z'$ mass,
$m_{Z'}$ = 0.5, 1, and $1.5~\rm{TeV}$ with $m_\chi = 200~\rm{GeV}$.
As seen in the figure, increasing $m_{Z'}$ moves the maxima of the distributions to higher values of $\xi$.
On the other hand, the background distribution has a peak at low $\xi$.
Therefore, we impose the lower cut on $\xi$ to reduce the SM background events at small $\xi$.
In the following analysis, we adopt the following selection cut on $\xi$:
\begin{align}
0.05 < \xi < 0.15,
\label{eq:xi_cut}
\end{align}
where the upper cut on $\xi$ is determined by the acceptance of the forward proton detectors in Eq.~\eqref{eq:xi_acceptance}.

\begin{table}[t!]\center
  \begin{tabular}{l|cccc}
    \hline
     &$N_B$&\multicolumn{3}{c}{$N_S: m_{Z'} = 1$~TeV, $m_\chi = 200$~GeV} \\
    Selection            &             & Vector & Axial & Mix\\
    \hline
     \begin{tabular}{l}
    minimal selections:\\ $p_T^{j}>200~{\rm GeV}, |\eta^{j}|<3.0$
     \end{tabular}&
    \begin{tabular}{c}8137\\ $(3.9$~fb)
    \end{tabular}&
    \begin{tabular}{c}
    579\\ $(0.28$~fb)
    \end{tabular}&
    \begin{tabular}{c}
    507\\ $(0.24$~fb)
    \end{tabular}&
     \begin{tabular}{c}
    546\\ $(0.26$~fb)
    \end{tabular}\\
   (a)(minimal selections)+$(0.015 < \xi < 0.15)$& 5027 & 389 & 340 & 365  \\
   (b)(minimal selections)+$(0.05 < \xi < 0.15)$& 2305 & 276 & 243  &258\\
    \hline
  \end{tabular}
  \caption{A cut-flow table for the signals and backgrounds
  at $\sqrt{s}=14~\mathrm{TeV}$ and $L_{\mathrm{int}}=3000~\mathrm{fb}^{-1}$.
The signal process is  $pp\to p\gamma p\to p j \chi \bar{\chi} X$ [Eq.~\eqref{eq:prod_proc}] and the 
background process is described by Eq.~\eqref{eq:bg_proc}.
The vector and axial-vector couplings of $Z'$ are given in Eqs.~(\ref{eq:lag_dm1})--(\ref{eq:lag_dm3}).
For the signals, $m_{Z'} = 1$~TeV and $m_\chi = 200$~GeV are selected. The survival probability $S^2$ is 0.7.
The cross sections of the minimal event selections are given in fb.}
\label{tab:cut-flow}
\end{table}
A cut flow is shown in Table~\ref{tab:cut-flow}.
From Table~\ref{tab:cut-flow}, for the cut condition (a),
the (square root of the) event numbers are reduced to
62\% of $N_B$ (79\% of $\sqrt{N_{B}}$) and 67\% of $N_S$ for the vector scenario,
where $N_{B}$ and $N_{S}$ are the number of background and signal events.
For the cut condition (b), the event numbers are reduced to
28\% of $N_B$ (53\% of $\sqrt{N_{B}}$) and 48\% of $N_S$  for the vector scenario.
 $N_{S}/\sqrt{N_{B}}$ (see Sec.~\ref{sec:results}) is slightly improved for selection cut (b) of Eq.~\eqref{eq:xi_cut}.

In our study, we do not include single diffractive productions or QCD processes.
Diffraction usually dominates for $\xi < 0.05$~\cite{Sun:2014ppa},
whereas the range of $\xi$ in our study is
$0.05 < \xi < 0.15$ [Eq.~\eqref{eq:xi_cut}].

It is known that
the pileup events are also significant backgrounds in addition to the SM process (\ref{eq:bg_proc}).
Although the forward proton is absent in $pp \to j \ensuremath{\slashed{E}_T} X$, the pileup events
can produce forward protons in the final state, which overlap with these hard scattering events.
For the forthcoming Run-III, the average number of pileup events per bunch crossing is assumed to be more than 50.
Even if there are 50 pileup events on average, there are always multiple protons in the forward region either
from diffractive production or inside the proton remnant in the case of inelastic scattering~\cite{Cho:2015dha}.
%
%
Therefore, it is hard to distinguish the process $pp\to j \ensuremath{\slashed{E}_T} X$ with one final forward proton (from pileup events)
from our process, i.e., $pp\to p\gamma p\to p j \ensuremath{\slashed{E}_T} X$.
Note that for each signal and background process,
the cross section of $pp\to j \ensuremath{\slashed{E}_T} X$ is about 1000 times larger than that of our
$pp\to p\gamma p\to p j \ensuremath{\slashed{E}_T} X$.
For example, the cross sections of processes mediated by $Z'$ are $\sigma(pp\to p\gamma p\to p j \ensuremath{\slashed{E}_T} X) \simeq 0.1$~fb and $\sigma(pp\to j \ensuremath{\slashed{E}_T} X) \simeq 0.2$~pb
for $(m_{Z'}, m_\chi) =(1.2~\mathrm{TeV}, 550~\mathrm{GeV})$ in the vector coupling scenario.
The corresponding SM background processes are $\sigma(pp\to p\gamma p\to p j \ensuremath{\slashed{E}_T} X)\simeq 4$~fb and
$\sigma(pp\to j \ensuremath{\slashed{E}_T} X)\simeq 12$~pb.
Therefore, unless the pileup events are controlled well enough,
the process $pp\to j \ensuremath{\slashed{E}_T} X$ with one final forward proton from pileup events
overwhelmingly dominates over our process.
In this case, $Z'$ and its mediated DM in $pp\to j \ensuremath{\slashed{E}_T} X$
should be the subject of an energetic-jet analysis~\cite{CMS:2017tbk,Aaboud:2017phn}.
In the next section we draw limit curves from the energetic-jet analysis~\cite{Aaboud:2017phn} in Fig.~\ref{fig:result}.
In our paper, we assume that the pileup events are sufficiently suppressed, and 
we leave an investigation of the suppression mechanism to the  future work.
For some ideas of how to separate the pileup events from the signal events,
see Refs.~\cite{Cho:2015dha,Baldenegro:2018hng,Harland-Lang:2018hmi,Tasevsky:2014cpa}.
The authors of Ref.~\cite{Cho:2015dha} used the fact that, for background events caused by pileup events, the forward proton and the particles in the main detector are produced by different proton-proton interactions.
The authors of Ref.~\cite{Tasevsky:2014cpa} discussed the possibility of
reducing pileup background events by measuring the time of flight of the deflected protons between the interaction point and the timing detectors.


A cut on the proton $p_T$ is also useful for reducing extra backgrounds, which occurs when the proton dissociates,
e.g., $p\to N^{*} + \gamma$ or $p\to \Delta +\gamma$
and decays back to a proton ~\cite{Harland-Lang:2018hmi}.

\section{Constraints using proton tagging at the forward proton detectors}\label{sec:results}
\begin{figure}[t]\center
\includegraphics[width=0.9\textwidth,clip]{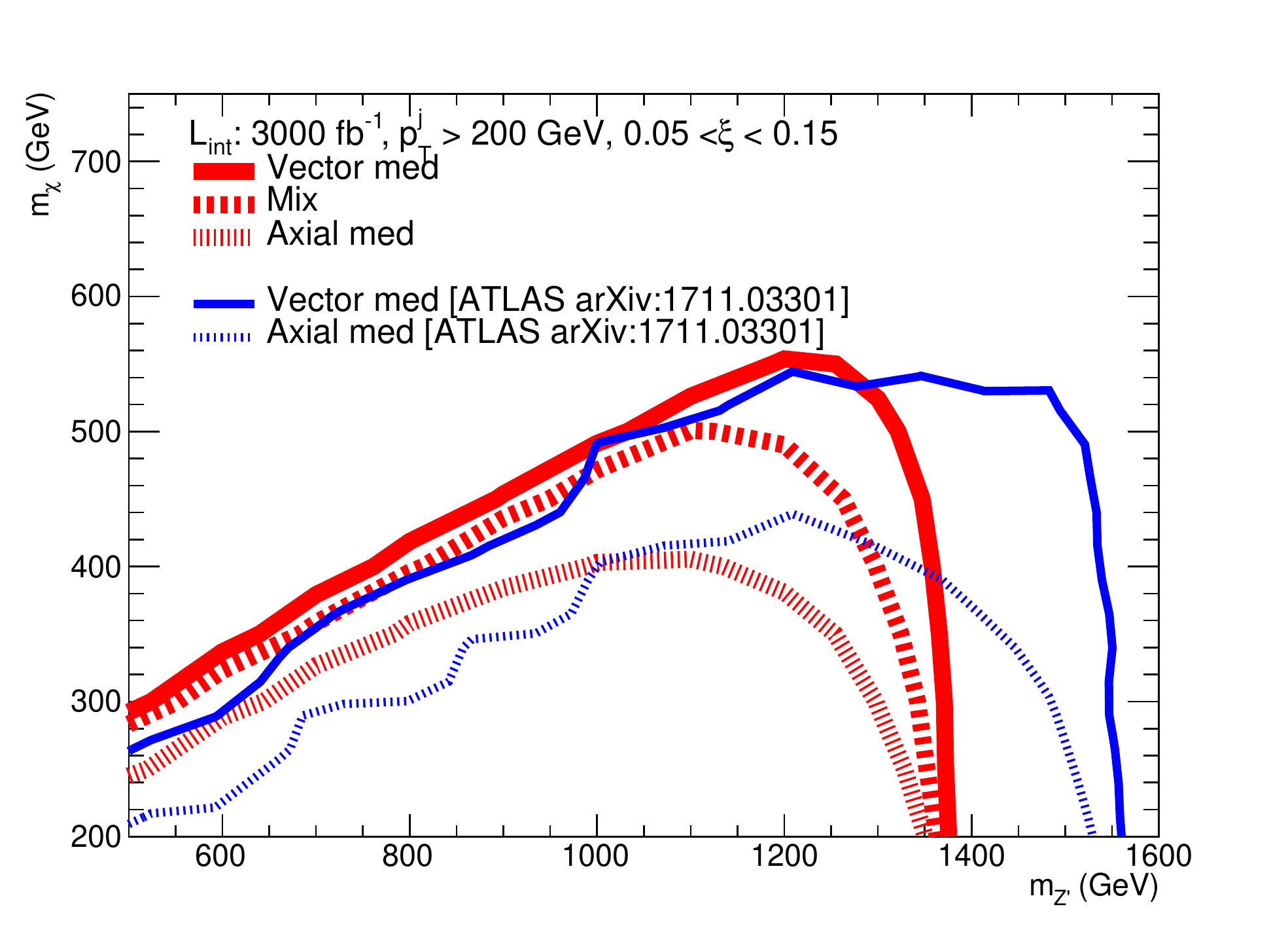}
\caption{The 95~\% C.L. exclusion limits in the $m_{Z'}$-$m_{\chi}$ plane.
The solid, dotted, and dashed lines correspond to
the vector scenario (\ref{eq:lag_dm1}),
the axial-vector scenario (\ref{eq:lag_dm2}), and
the mixed scenario (\ref{eq:lag_dm3}), respectively.
The red thick curves are for our photon-induced processes using forward detectors and the blue thin curves are
from an energetic-jet analysis from $pp$ collision (observed limits)~\cite{Aaboud:2017phn}.
}
\label{fig:result}
\end{figure}
In this section we investigate constraints on the parameter space of the simplified DM model with a leptophobic spin-1 mediator using proton tagging at the forward proton detector.
The exclusion limits on the model parameters ($m_{Z'}, m_\chi$) are imposed by assuming a null observation.
We calculate $N_{S}/\sqrt{N_{B}}$ by scanning over the mediator mass $m_{Z'}$ and the DM mass $m_{\chi}$. 
The lower limits on $m_{Z'}$ and $m_\chi$ at the 95~\% confidence level (C.L.) are determined by requiring $N_{S}/\sqrt{N_{B}} > 1.96$.
After the selection cuts on the kinematical variables shown in Sec.~\ref{sec:analysis},
we find $N_B =  2305$ for $\sqrt{s}=14~\mathrm{TeV}$  with an integrated luminosity $L_{\rm{int}} = 3000~\rm{fb}^{-1}$.
This event number allows for a reasonable estimate of the statistical uncertainty.

We show the exclusion limits on the model parameters $(m_{Z'}, m_\chi)$
in Fig.~\ref{fig:result} for the three scenarios of the interaction of $Z'$ in Sec.~\ref{sec:model}.
The solid, dotted, and dashed red thick curves represent the limits for the vector scenario (\ref{eq:lag_dm1}), the axial-vector scenario (\ref{eq:lag_dm2}),
and the mixed scenario (\ref{eq:lag_dm3}), respectively.
We also draw the limits for the vector scenario
and the axial-vector scenario from an energetic-jet analysis from $pp$ collisions (observed limits)~\cite{Aaboud:2017phn},
shown by the thin-solid and thin-dotted blue curves, respectively.
Note that the mixed scenario was not studied in their analysis.
It can be seen from the figure that
the limit on $(m_{Z'}, m_\chi)$ goes from weakest to strongest for scenarios
 (\ref{eq:lag_dm1}), (\ref{eq:lag_dm3}), and (\ref{eq:lag_dm2}).
This result is consistent with the signal event number distributions in Sec.~\ref{sec:analysis}.
The figure also tells us that
the mediator mass $m_{Z'} \lesssim 1.4~\mathrm{TeV}$ is excluded at 95\% C.L. for all scenarios when the DM mass is relatively small.
%
%
On the other hand, the lower mass bound on the DM $\chi$ can be summarized as follows.
\begin{itemize}
  \item The vector scenario (\ref{eq:lag_dm1}):
    \begin{align}
    m_\chi \gtrsim 550~\mathrm{GeV}~~(m_{Z'} = 1.2~\mathrm{TeV}).
    \label{lim_mchi1}
  \end{align}
  \item The axial-vector scenario (\ref{eq:lag_dm2}):
  \begin{align}
    m_\chi \gtrsim 400~\mathrm{GeV}~~(m_{Z'} = 1.1~\mathrm{TeV}).
    \label{lim_mchi2}
  \end{align}
  \item The mixed scenario (\ref{eq:lag_dm3}):
  \begin{align}
    m_\chi \gtrsim 500~\mathrm{GeV}~~(m_{Z'} = 1.1~\mathrm{TeV}).
    \label{lim_mchi3}
  \end{align}
\end{itemize}

The limits on the DM mass for the vector scenario (\ref{lim_mchi1}) 
and the axial-vector scenario (\ref{lim_mchi2}) in Fig.~\ref{fig:result}
are slightly stronger than the energetic-jet analysis from the ATLAS study~\cite{Aaboud:2017phn} for
$m_{Z'} \lesssim 1.2$~TeV (the vector scenario) and 
$m_{Z'} \lesssim 1$~TeV (the axial-vector scenario), respectively.
The limits on the DM mass for each scenario are mainly weaker than those from the ATLAS study for the other parameter space,
i.e., $m_{Z'} \gtrsim 1.2$~TeV for the vector scenario and $m_{Z'} \gtrsim 1$~TeV for the axial-vector scenario.
On the other hand,
the limit on the DM mass in the mixed scenario  (\ref{lim_mchi3}) 
has not been studied at the LHC.
The recent combined result of dijet invariant-mass searches at ATLAS and CMS gives the lower mass bound on $Z'$
as $m_{Z'} \gtrsim 5~\mathrm{TeV}$~\cite{Bagnaschi:2019djj}
with the observed limit value, which is much stronger than our result on $m_{Z'}$.

Throughout this paper we have not discussed any astrophysical
constraints on the simplified DM models.
For constraints on the model parameter space from these types of observations or experiments,
see Refs.~\cite{CMS:2017tbk,DEramo:2016gos,Ellis:2018xal,Bagnaschi:2019djj}.
The relic density constraints from the Planck satellite experiment show a strong
limit on the parameter space for the axial-vector scenario  (\ref{lim_mchi2}),
whereas the direct--detection constraints give severe limits on the parameter space for the
vector scenario  (\ref{lim_mchi1})~\cite{CMS:2017tbk}.
Indirect--detection constraints on the simplified DM models
are considered as unimportant when the DM mass $m_\chi \gtrsim 50$~GeV~\cite{Ellis:2018xal,Bagnaschi:2019djj}.
\section{Summary}\label{sec:summary}

We have studied the feasibility of searching for a
simplified DM model with a leptophobic vector-mediator
using the forward proton detectors at the LHC.
In our study
we investigated the fermionic DM $\chi$ production process
$pp \to p\gamma p \to p j \chi \bar{\chi}X$, based on three
scenarios for the interactions of the mediator $Z'$ with quarks $q$ or the DM $\chi$.
In the first scenario (``vector scenario"), $Z'$ couples to quarks $q$ and $\chi$ through vector couplings $g^V_q$ and $g^V_\chi$, respectively.
In the second scenario (``axial-vector scenario"),
these vector-type interactions are replaced by axial-vector interactions with couplings $g^A_q$ and $g^A_\chi$.
The third scenario  (``mixed scenario") uses
both vector and axial-vector couplings.
Our study was performed at the parton level.
%

We found that the selection cut on $\xi$, which is defined as the momentum fraction loss of
intact protons detected at the forward proton detectors, is very useful for reducing the background events.
In our study, we focused on the main SM background process $pp\to p\gamma p\to p j \nu \bar{\nu} X$. 
We did not take into account the effect of pileup events for both signal and backgrounds in our analysis for the constraints 
on the model parameter space.
Taking account of event selection conditions,
constraints on the model parameter space at the LHC were obtained for
$\sqrt{s}=14~\mathrm{TeV}$ and an integrated luminosity $L_{\mathrm{int}}=3000~\mathrm{fb}^{-1}$.
The lower bound on the mediator mass $m_{Z'}$ at the 95\% C.L, is about $1.4~\mathrm{TeV}$ and no significant difference in the lower bound among the three scenarios was found.
The lower limit on the DM mass at
95\% C.L. is $m_\chi \gtrsim 550~\mathrm{GeV}$ at $m_{Z'}=1.2~\mathrm{TeV}$ for the vector scenario,
$m_\chi \gtrsim 400~\mathrm{GeV}$ at $m_{Z'}=1.1~\mathrm{TeV}$ for the axial-vector scenario,
and
$m_\chi \gtrsim 500~\mathrm{GeV}$ at $m_{Z'} = 1.1~\mathrm{TeV}$ for the mixed scenario.

The processes that we studied in this paper are not the conventional QCD processes at the LHC. 
The forward proton detectors at the LHC provide us with opportunities to test new photon-induced processes and might give a chance to look for physics beyond the SM.
%



\paragraph*{Acknowledgements}
We thank Kentarou Mawatari, Cen Zhang, and Chen Zhang for their valuable comments and discussions.
K.Y. is supported by the Chinese Academy of Sciences (CAS) President's International Fellowship Initiative under Grant No. 2020PM0018.
K.Y.'s work was also supported in part by the National Center for Theoretical Sciences, Taiwan.
The work of G.C.C.
is supported in part by Grants-in-Aid for Scientific Research from the Japan Society for the Promotion of Science (No. 16K05314).

\end{document}